\documentclass[12pt]{article}
\usepackage{fullpage,amsmath}
\usepackage[pdftex,dvipsnames,usenames]{color}         %for colors
\usepackage{xcolor}
\usepackage{natbib,bm}
\usepackage{float}
\usepackage{lscape}
\usepackage{bbm}
\usepackage{amsthm} %to include command ``\newtheorem*`` 
\newcommand{\comment}[1]{}

\newcommand{\nmathbf}{\bm}

\def\bfA{\nmathbf A}

\def\bfD{\nmathbf D}

\def\bfS{\nmathbf S}

\def\bfU{\nmathbf U}
\def\bfV{\nmathbf V}
\def\bfW{\nmathbf W}

\def\bfZ{\nmathbf Z}

\def\bfd{\nmathbf d}

\def\bfp{\nmathbf p}

\def\bfs{\nmathbf s}

\def\bfy{\nmathbf y}

\def\bfalpha  {\nmathbf \alpha}
\def\bfbeta   {\nmathbf \beta}

\def\bfmu     {\nmathbf \mu}

\def\bfrho    {\nmathbf \rho}

\def\bfSigma  {\nmathbf \Sigma}

\def\boldfacefake#1{\kern-4pt
   \hbox{ \mathsurround=0pt
   \hbox to 0.4pt{$#1$\hss}\hbox to 0.4pt{$#1$\hss}\hbox {$#1$}}}

\newcommand{\Cov}{\mbox{Cov}}
\newcommand{\Var}{\mbox{Var}}

\newcommand{\ba}{\begin{eqnarray*}}
\newcommand{\ea}{\end{eqnarray*}}

\newtheorem{theorem0}{Theorem}
\newtheorem{lemma0}{Lemma}
\newtheorem{remark0}{Remark}

\newtheorem{example0}{Example}
\newtheorem{definition0}{Definition}
\newtheorem{corollary0}{Corollary}
\newtheorem{proposition0}{Proposition}
\newtheorem{algorithmY}{Algorithm}
\newtheorem{condition0}{Condition}
\newtheorem{assumption0}{Assumption}
\newtheorem{simulation0}{Simulation}

\newcommand{\reals}{\mbox{\rm I\kern-.20em R}}
\newcommand{\sreals}{\mbox{\small \rm I\kern-.20em R}}

\ifx\pdfoutput\undefined 
\usepackage[dvips]{graphicx}
\else
\usepackage[pdftex]{graphicx}

\usepackage{epstopdf}

\newcommand{\bqg}{\begin{quote} \color{Green}\em}%X
	\newcommand{\toale}{\end{quote} \color{black}\rm}
\newcommand{\eqg}{\end{quote} \color{black}\rm}%X
\newcommand{\bd}{\begin{description}}
\newcommand{\ed}{\end{description}}
\newcommand{\bi}{\begin{itemize}}
\newcommand{\ei}{\end{itemize}}
\newcommand{\be}{\begin{enumerate}}
\newcommand{\ee}{\end{enumerate}}

\begin{document}

	\begin{center}
		{\large \bf Sample Size Calculation for Cluster Randomized Trials with Zero-inflated Count Outcomes
		}
		
		\vspace{.5 cm}
		
		Zhengyang Zhou, Ph.D. \\
		University of North Texas Health Science Center, Fort Worth, TX \\
		
		Dateng Li, Ph.D.\footnote{Correspondence should be sent to: Dateng Li, Ph.D. Email:  dateng.li@regeneron.com} \\
		Regeneron Pharmaceuticals Inc., Tarrytown, NY \\

		Song Zhang, Ph.D.  \\
		University of Texas Southwestern Medical Center, Dallas, TX\\
		
	\end{center}
	\vspace{1 cm}
	\begin{abstract}
Cluster randomized trails (CRT) have been widely employed in medical and public health research. Many clinical count outcomes, such as the number of falls in nursing homes, exhibit excessive zero values. In the presence of zero inflation, traditional power analysis methods for count data based on Poisson or negative binomial distribution may be inadequate. In this study, we present a sample size method for CRTs with zero-inflated count outcomes. It is developed based on GEE regression directly modeling the marginal mean of a ZIP outcome, which avoids the challenge of testing two intervention effects under traditional modeling approaches. A closed-form sample size formula is derived which properly accounts for zero inflation, ICCs due to clustering, unbalanced randomization, and variability in cluster size.  Robust approaches, including \textit{t}-distribution-based approximation and Jackknife re-sampling variance estimator, are employed to enhance trial properties under small sample sizes. Extensive simulations are conducted to evaluate the performance of the proposed method. An application example is presented in a real clinical trial setting.
	\end{abstract}
	\vspace{1 cm}
\section{Introduction}
Clinical trials that perform randomization at the cluster level (e.g., clinics, schools, communities, etc.) have been widely used to assess effectiveness of interventions in medical and public health research. Such trials are commonly referred to as cluster randomized trials (CRTs; \cite{murray2004design, eldridge2012practical}). Since participants in the same cluster share certain characteristics (e.g., the same physician, teacher, similar socioeconomic status, etc), their responses tend to be positively correlated. This intracluster correlation coefficient (ICC) is one of the key features that needs to be considered in sample size calculation for CRTs \citep{murray1998design}. 

Count outcomes are frequently used in RCTs. Examples include number of cigarettes in a smoking cessation study \citep{roig2010cluster}, number of clinic visits in a educational outreach study \citep{fairall2005effect}, and number of days in ICU in a nutritional support study \citep{martin2004multicentre}. To model count data, the Poisson distribution has been widely used \citep{cameron2013regression,agresti2003categorical}. Many researchers, however, have reported the phenomenon of zero inflation, where the observed proportion of zeros is much greater than the theoretical proportion under Poisson \citep{lambert1992zero, lewsey2004utility, moghimbeigi2008multilevel}. For example, in a long-term care (LTC) RCT evaluating the effect of multifaceted knowledge translation for care teams, one outcome was the number of falls over three months among senior residents in LTC homes, which exhibited substantial zero-inflation with more than 30\% residents did not have any fall in both control and intervention groups during the follow-up period \citep{kennedy2015successful}. Furthermore, zero-inflated count data usually manifests overdispersion \citep{yang2009testing}, which violates the assumption of mean and variance being equal under the Poisson model. Imposing the Poisson assumption on zero-inflated count outcomes leads to poor statistical results including biased estimation and under-estimated sample sizes.  Recently, the negative binomial (NB) model has gained popularity due to its flexibility in accommodating overdispersion \citep{cameron2013regression,agresti2003categorical}. The NB model, however, does not address the issue of zero-inflation either.

Most of existing sample size methods for CRTs with count outcomes are developed under the Poisson model. For example, \cite{amatya2013} proposed a sample size calculation method based on Poisson regression. It required the assumption of equal cluster size, which might be unrealistic in real-world clinical settings. It has been shown that ignoring variability in cluster size leads to under-powered studies \citep{ahn2014sample, liu2018relative}.  \cite{wang2018} relaxed this assumption and proposed a sample size method accommodating randomly varying cluster sizes. It included a correction term involving a coefficient of variance for cluster sizes. \cite{li2019sample} proposed a sample size method for correlated count outcome based on the NB model. Developed based on either Poisson or NB, the aforementioned methods are inapplicable to CRTs where the count outcomes contain excessive zeros. To the best of our knowledge, there has been no investigation on sample size calculation for zero-inflated count data in CRTs.

The zero-inflated Poisson (ZIP) model has been widely used to analyze count data with excessive zeros \citep{mullahy1986specification,preisser2012review,famoye2006zero}. It assumes the count data to arise from a mixture of a Poisson distribution and a point mass at zero (i.e., the structural zero). This mixture distribution is characterized by two parameters: the Poisson mean and the probability of structural zero. Separate hypothesis testings can be performed on these two parameters, which assess the intervention effect in two dimensions. %: the Poisson mean (e.g., incidence rate ratio) and the structural zero rate (e.g., odds ratio). 
As a result, sample size calculation based on a ZIP model is conceptually challenging due to the need of simultaneously testing two hypotheses. \cite{long2014marginalized} presented an alternative approach. Instead of evaluating the two intervention effects, they proposed to directly make inference on the overall intervention effect, quantified by the marginal mean under the ZIP framework. This approach is denoted as the marginalized ZIP model. In this study, we build upon this idea to develop power analysis methods for the comparison of overall intervention effect in CRTs with zero-inflated count outcomes. The resulting sample size formula has a closed form, which facilitates implementation and enable researchers to analytically assess the impact of various design parameters. It also accommodates pragmatic design issues frequently encountered by practitioners such as unbalanced randomization and randomly varying cluster sizes. 

The rest of the paper is organized as follows. In Section 2
we describe the statistical model and power analysis approach
for CRTs with zero-inflated outcomes. In Section 3 we conduct extensive simulations to evaluate the performance of the proposed method. In Section 4 a real application example is presented. In Section 5 we provide discussion and concluding remarks.

\section{Methodology}
\subsection{Statistical model and sample size}
Suppose $N$ clusters are randomized to the control or intervention arm in a
CRT.
We use $m_i$ $(i=1,...,N)$ to denote the cluster sizes and $m_i$ are assumed to follow a certain discrete distribution: $\mbox{Prob}(m_i=m)=g(m)$
with outcome space $\mathcal{M}$. We define mean $\eta_m = E(m_i)$ and variance
$\sigma_m^2 = \Var(m_i)$. Let $y_{ij}$ be the count outcome measured on the
$j$th subject from the $i$th cluster. We assume that $y_{ij}$ arises from a ZIP distribution, which is the mixture of two components: a point mass at zero with probability $p_{ij}$, and a Poisson distribution of mean $\lambda_{ij}$
with probability $1-p_{ij}$. Presented through latent variables, we have 
\begin{eqnarray*}
y_{ij}=\left\{\begin{array}{l l}
      0&~ \mbox{if}~s_{ij}=1 ;  \\
      u_{ij}&~\mbox{if}~s_{ij}=0,
\end{array}
\right. 
\end{eqnarray*}
where $s_{ij}$ is binary with $\mbox{Prob}(s_{ij}=1)=p_{ij}$ and $u_{ij} \sim \mbox{Poisson}(\lambda_{ij})$. Here $s_{ij}$ and $u_{ij}$ are latent variables. We define ICCs $\rho_s=\mbox{Corr}(s_{ij}, s_{ij'})$ and $\rho_u=\mbox{Corr}(u_{ij}, u_{ij'})$ for $j\neq j'$. Responses are assumed to be independent across clusters. It is straightforward that Prob$(y_{ij}=0)=e^{-\lambda_{ij}}+p_{ij}$, with $p_{ij}$ accounting for the extra zeros than those expected from a Poisson($\lambda_{ij}$). \cite{beckett2014zero} show that the marginal mean and variance of $y_{ij}$ are
\begin{equation}\label{eqn_mu}
E(y_{ij}) = \mu_{ij} = (1-p_{ij})\lambda_{ij}
\end{equation}
and
\begin{equation}\label{eqn_var_y}
\Var(y_{ij}) =
\mu_{ij}+\frac{p_{ij}}{1-p_{ij}}\mu_{ij}^2.
\end{equation}
It is obvious that $\Var(y_{ij})$ is an increasing function of $\mu_{ij}$ and $p_{ij}$. Furthermore, $\Var(y_{ij})>E(y_{ij})$ always holds. That is, given the same mean, a ZIP variable has a larger variance than a Poisson variable. Hence zero inflation leads to over-dispersion. Mis-specifying a Poisson model for a ZIP variable would lead to under estimated variability in data analysis, and under-powered clinical trials in experimental design. The severity of over-dispersion is associated with a larger mean ($\mu_{ij}$) and a larger probability of structural zero ($p_{ij}$).  

Traditionally researches have evaluated the intervention effect by testing two hypotheses, one constructed based on $p_{ij}$ and the other based on $\lambda_{ij}$. Such approaches lead to difficulty in sample size calculation because statistical inference involves testing two hypotheses. For CRTs, this difficulty is further complicated by the need to consider clustering. 

In this study we propose to directly evaluate the overall intervention effect, measured on the marginal mean ($\mu_{ij}$) of a ZIP outcome. Specifically, we assume  
\begin{equation}\label{overall_model}
\log(\mu_{ij}) = \beta_1 + \beta_2r_i.
\end{equation}
Here $r_i=0/1$ indicates that the $i$th cluster is randomized to the
control/intervention arm and $\beta_{2}$ is the difference in marginal mean between the intervention
and control group on the log scale, representing the overall intervention
effect \citep{long2014marginalized}. A cluster receives intervention with probability $\bar{r} = E(r_i)$. Define $\mu_i=\exp(\beta_1+\beta_2r_i)$ and Model (\ref{overall_model}) suggests that $\mu_{ij}=\mu_i$. Similarly, we assume $p_{ij}=p_i$ and $\lambda_{ij}=\lambda_i=\mu_i/(1-p_i)$.  The hypotheses of interest are $H_0:\beta_2=0$ vs $H_1:\beta_2 \neq 0$.

%In this study we propose to take an alternative approach, constructing models for $p_{ij}$ and $\mu_{ij}$. Sample size calculation is then developed for the inference about treatment effect on $\mu_{ij}$, the marginal mean, which summarizes the treatment effects on both components. Specifically, a logit model is assumed for $p_{ij}$, 
%\begin{equation}\label{zero-degenerated}
%\log\left(\frac{p_{ij}}{1-p_{ij}}\right) = \alpha_1 + \alpha_2r_i,
%\end{equation}
%where $r_i=0/1$ indicates that the $i$th cluster is randomized to the
%control/intervention arm and $\alpha_2$ represents treatment effect on the probability of structural zeros. A cluster receives the intervention with probability $\bar{r} = E(r_i)$. 

With Equations (\ref{overall_model}) and (\ref{eqn_var_y}), models for the first two moments of $y_{ij}$ have been specified. We can estimate the regression parameters $\bfbeta$ using the generalized estimating equation (GEE) approach \citep{liang1986}. Define $\bfZ_{ij} = (1,r_i)'=\bfZ_i$ and $\bfbeta = (\beta_1, \beta_2)'$. Let $\bfy_i=(y_{i1}, \cdots, y_{im_{i}})'$ be the cluster-specific response vector with mean 
$\bfmu_{i}(\bfbeta)=[\mu_{i1}(\bfbeta), \cdots,
\mu_{im_i}(\bfbeta)]' = \mu_i\bm{ \mathbbm{1}}_{m_i}$, where $\mu_i=\exp(\bfZ_i'\bfbeta)$ and $\bm{ \mathbbm{1}}_{m_i}$ is a vector of length $m_i$ with all elements being $1$. Utilizing the independent working correlation, the GEE estimator $\hat{\bfbeta}=(\hat{\beta}_{1},\hat{\beta}_{2})'$ is the solution to score function  
\begin{equation}\label{eqn_GEE}
S_{N}(\bfbeta)=N^{-\frac{1}{2}}\sum_{i=1}^{N} \bfD_i' \bfW_i^{-1} [\bfy_i -
\boldsymbol{\bfmu_i}(\bfbeta) ]= \bf0,
\end{equation}
where $\bfD_i =\frac{\partial \boldsymbol{\bfmu_i}(\bfbeta)}{\partial \bfbeta}$ is an $m_i\times 2$ gradient matrix and $\bfW_i$ is an $m_i\times m_i$ diagonal matrix with all diagonal elements being 
$\mu_{i}+\frac{p_{i}}{1-p_{i}}\mu_{i}^2$. Equation (\ref{eqn_GEE}) can be solved through the Newton-Raphson algorithm.
Specifically, at the $l$th iteration, 
\begin{equation}\label{N-R}
\hat{\bfbeta}^{(l)} = \hat{\bfbeta}^{(l-1)} +
N^{-\frac12}\bfA_{N}^{-1}(\hat{\bfbeta}^{(l-1)})\bfS_{N}( \hat{\bfbeta}^{(l-1)}), 
\end{equation}
where
\begin{equation}\label{An}
\bfA_{N}(\hat{\bfbeta})= N^{-1}\sum_{i=1}^{N}\sum_{j=1}^{m_i} \bfZ_{i}\bfZ'_{i}\frac{\mu_{i}
	(\hat{\bfbeta})}{1 + \frac{p_i}{1-p_i}\mu_{i} (\hat{\bfbeta})}.
\end{equation}
%The computation in (\ref{N-R}) requires the values of $\bfalpha=(\alpha_1,\alpha_2)'$. Researchers usually  assume $p_i$, or equivalently $\bfalpha=(\alpha_1, \alpha_2)'$, to be known at the design stage. We will discuss the estimation of $\bfalpha$ later. 

As shown by \cite{liang1986}, $\sqrt{N}(\hat{\bfbeta}-\bfbeta)$
approximately follows a normal distribution with mean $\bf0$ and variance
$\bfSigma_{N}=\bfA_{N}^{-1}\bfV_{N}\bfA_{N}^{-1}$, where 
\begin{equation*}
\bfV_{N}(\hat{\bfbeta})=N^{-1}\sum_{i=1}^{N}\sum_{j=1}^{m_i}\sum_{j'=1}^{m_i}\hat{\epsilon}_{ij}\hat{\epsilon}_{ij'}\frac{1}{[1
	+ \frac{p_i}{1-p_i}\mu_{i} (\hat{\bfbeta})]^2}\bfZ_{i}\bfZ'_{i}.
\end{equation*}
Here $\hat{\epsilon}_{ij}=y_{ij}-\exp(\bfZ_{I}'\hat{\bfbeta})$ is the residual. Let $\hat{\sigma}_{2}^{2}$ 
be the (2,2)th element of $\bfSigma_{N}$. We reject $H_{0}:\beta_{2}=0$
if $\sqrt{n}|\hat{\beta}_{2}|/\hat{\sigma}_{2}>z_{1-\alpha/2}$, where $z_{1-\alpha/2}$ is the $100(1-\alpha/2)$th
percentile of the standard normal distribution. Define $\bfA$ and $\bfV$ to be the limits of $\bfA_{N}$ and $\bfV_{n}$ as $N\rightarrow\infty$. It
follows that
$\bfSigma_{N}$ converges to $\bfSigma=\bfA^{-1}\bfV\bfA^{-1}$. Let
$\sigma_{2}^{2}$ be the (2,2)th element of $\bfSigma$. Given the true intervention effect
$\beta_{2}=\beta_{20},$ the number of clusters to achieve power $1-\gamma$ at two-sided
type I error  $\alpha$ is calculated by
\begin{equation}\label{ss_eqn1}
N=\frac{\sigma_{2}^{2}(z_{1-\alpha/2}+z_{1-\gamma})^{2}}{\beta_{20}^{2}}.
\end{equation}
In the following we show that a closed-form expression of $\sigma_2^2$ can be derived, which
leads to a closed-form sample size formula. First, as $N\rightarrow\infty$, it is easy to show that $\bfA_{N}$ approaches
\begin{eqnarray}\label{A}
\bfA =(1-\bar{r})\frac{\mu^*_1\eta_m}{1+\frac{p_1^*}{1-p_1^*}\mu^*_1}\left(\begin{array}{cc}
1 & 0\\
0 & 0
\end{array}\right)
+\bar{r}\frac{\mu^*_2\eta_m}{1+\frac{p_2^*}{1-p_2^*}\mu^*_2}\left(\begin{array}{cc}
1 & 1\\
1 & 1
\end{array}\right).
\end{eqnarray}
Here we define $p_i=p^*_1$ to be the probability of structural zero under control ($r_i=0$) and $p_i=p^*_2$ under intervention ($r_i=1$). Similarly, $\mu^*_1 = \exp(\beta_1)$ and $\mu^*_2 = \exp(\beta_1 + \beta_2)$ are the marginal means. Recall that $\eta_m=E(m_i)$. As $N\rightarrow\infty$,  $\bfV_{N}(\hat{\bfbeta})$ approaches
\begin{eqnarray}
\bfV &=&
E\left[\sum_{j=1}^{m_i}\sum_{j'=1}^{m_i}\frac{[y_{ij}-\mu_{i}(\bfbeta)][y_{ij'}-\mu_{i}(\bfbeta)]}{[1
	+ \frac{p_i}{1-p_i}\mu_{i} (\bfbeta)]^2}
\left(\begin{array}{cc}
1 & r_{i}\\
r_i & r_i^2
\end{array}\right)\right].\\\nonumber
&=&(1-\bar{r})\bfV_1
+\bar{r}\bfV_2,
\end{eqnarray}
where
\begin{eqnarray}
\bfV_1 &=&  \left(\begin{array}{cc}
1 & 0\\\nonumber
0 & 0

\end{array}\right)\left\{\eta_m\frac{\mu^*_1+\frac{p^*_1}{1-p^*_1}\mu_1^{*2}}{[1+\frac{p^*_1}{1-p^*_1}\mu^*_1]^2}
+ (\eta_m^2+\sigma_m^2-\eta_m)\frac{\zeta_1}{[1+\frac{p^*_1}{1-p^*_1}\mu^*_1]^2}\right\},
\end{eqnarray}
and 
\begin{eqnarray}
\bfV_2 &=&  \left(\begin{array}{cc}
1 & 1\\\nonumber
1 & 1
\end{array}\right)\left\{\eta_m\frac{\mu^*_2+\frac{p^*_2}{1-p^*_2}\mu_2^{*2}}{[1+\frac{p^*_2}{1-p^*_2}\mu^*_2]^2}
+ (\eta_m^2+\sigma_m^2-\eta_m)\frac{\zeta_2}{[1+\frac{p^*_2}{1-p^*_2}\mu^*_2]^2}\right\}.
\end{eqnarray}
The terms $\zeta_1$ and $\zeta_2$ have relatively complicated expressions,
\begin{equation*}
\zeta_1 =
\mu^*_1\left[\frac{p^*_1}{1-p^*_1}\rho_s-2(\mu^*_1+2)p_1^{*2}(\rho_s-1)+
p^*_1(\rho_s-1)(\rho_u + 2)+\rho_u\right],
\end{equation*}
and
\begin{equation*}
\zeta_2 =
\mu^*_2\left[\frac{p^*_2}{1-p^*_2}\rho_s-2(\mu^*_2+2)p_2^{*2}(\rho_s-1)+
p^*_2(\rho_s-1)(\rho_u + 2)+\rho_u\right].
\end{equation*}
Derivation details of $\bfV_1$ and $\bfV_2$ are presented in Appendix A. 

Using matrix algebra, we can obtain the (2,2)th element of
$\bfSigma=\bfA^{-1}\bfV\bfA^{-1}$: 
\begin{equation}\label{sigma2sq}
\sigma_2^2 =
\frac{\eta_m\mu^*_1(1+\frac{p^*_1}{1-p^*_1}\mu^*_1)+(\eta_m^2+\sigma_m^2-\eta_m)\zeta_1}{(1-\bar{r})\mu_1^{*2}\eta_m^2}+
\frac{\eta_m\mu^*_2(1+\frac{p^*_2}{1-p^*_2}\mu^*_2)+(\eta_m^2+\sigma_m^2-\eta_m)\zeta_2}{\bar{r}\mu_2^{*2}\eta_m^2}.
\end{equation}
Plugging (\ref{sigma2sq}) into Equation (\ref{ss_eqn1}) gives the closed-form sample size formula:
\begin{equation}\label{ss_eqn2}
N^{(z)}=\frac{\left[ \frac{\eta_m\mu^*_1(1+\frac{p^*_1}{1-p^*_1}\mu^*_1)+(\eta_m^2+\sigma_m^2-\eta_m)\zeta_1}{(1-\bar{r})\mu_1^{*2}\eta_m^2}+
\frac{\eta_m\mu^*_2(1+\frac{p^*_2}{1-p^*_2}\mu^*_2)+(\eta_m^2+\sigma_m^2-\eta_m)\zeta_2}{\bar{r}\mu_2^{*2}\eta_m^2} \right](z_{1-\alpha/2}+z_{1-\gamma})^{2}}{\beta_{20}^{2}}.
\end{equation}
Sample size $N^{(z)}$ is obtained under asymptotic normal approximation. In practice, when the number of clusters is limited, the normal approximation might not perform well. In such cases, an alternative approach is to use the $t$-distribution. Usually there is no closed-form formula for sample size calculation based on the  $t$-distribution. \cite{tang2017closed}  proposed a two-step procedure to obtain sample size under the $t$-distribution:
\begin{equation}\label{ss_eqn3}
N^{(t)}=\frac{\left[ \frac{\eta_m\mu^*_1(1+\frac{p^*_1}{1-p^*_1}\mu_1)+(\eta_m^2+\sigma_m^2-\eta_m)\zeta_1}{(1-\bar{r})\mu_1^{*2}\eta_m^2}+
\frac{\eta_m\mu^*_2(1+\frac{p^*_2}{1-p^*_2}\mu_2)+(\eta_m^2+\sigma_m^2-\eta_m)\zeta_2}{\bar{r}\mu_2^{*2}\eta_m^2}\right](t_{f(N^{(z)}),1-\alpha/2}+t_{f(N^{(z)}),1-\gamma})^{2}}{\beta_{20}^{2}}.
\end{equation}
Here the degree of freedom for the $t$-distribution is computed by a function of $N^{(z)}$, denoted as $f(N^{(z)})$. We set $f(N^{(z)})=N^{(z)}-2$, which equals to the number of clusters minus the number of regression parameters. 

In summary, to compute a sample size using (\ref{ss_eqn2}) or (\ref{ss_eqn3}), we need to specify the mean and variance of cluster sizes ($\eta_m,\sigma_m^2$), the randomization probability $\Bar{r}$, the regression parameters $\bfbeta$, the probabilities of structural zeros ($p_1^*, p_2^*$), the ICC parameters $(\rho_s,\rho_u)$, and pre-determined levels of tyep I error $\alpha$ and power $1-\gamma$. 

\subsection{Decomposing the marginal treatment effect $\beta_2$}
Following the definition of $(p^*_1, p^*_2)$ and $(\mu^*_1, \mu^*_2)$, we define $\lambda^*_1$ and $\lambda^*_2$ to be the Poisson mean under control and intervention, respectively. Hence $\mu^*_k=(1-p^*_k)\lambda^*_k$ for $k=1,2$. From  (\ref{overall_model}) we have
\begin{equation*}
\beta_2 =\log(\mu^*_2)-\log(\mu^*_1)=\log(\lambda^*_2)-\log(\lambda^*_1) + \log(1-p^*_2)-\log(1-p^*_1).
\end{equation*} 
That is, the overall intervention effect $\beta_2$ can be decomposed into $\log(\lambda^*_2)-\log(\lambda^*_1)$ and $\log(1-p^*_2)-\log(1-p^*_1)$, representing the effects on the Poisson part and the structural zero part, respectively. We introduce a new parameter $q$ such that 
\begin{equation}
\label{eq:q}
\log(1-p^*_2)-\log(1-p^*_1) = q\beta_2.
\end{equation} 
In practice the intervention usually affects the Poisson part and the structural zero part in the same direction. For example,  an intervention aimed at controlling alcohol consumption tends to increase the proportion of abstiners as well. Hence we assume that $\log(\lambda^*_2)-\log(\lambda^*_1)$ and $\log(1-p^*_2)-\log(1-p^*_1)$ are of the same sign and $q\in [0, 1]$. We interpret $q$ as the proportion of treatment effect due to change in the probability of structural zeros.  
%Furthermore, it can be shown that given $(\alpha_1,\beta_2,q)$, $\alpha_2$ is determined:
%\begin{equation}\label{alpha2_q_eqn}
%\alpha_2 = \log[\exp\{\log(1+e^{\alpha_1})-q\beta_2\}-1]-\alpha_1.
%\end{equation}

It is straightforward that $p^*_2=1-\exp(q\beta_2)(1-p^*_1)$. Hence during sample size calculation, we can equivalently specify either $(\beta_1,\beta_2, p^*_1, p^*_2)$, or $(\beta_1,\beta_2,p^*_1,q)$. We prefer the latter because it offers a straightforward decomposition of the overall intervention effect and a natural framework for sensitivity analysis. At the design stage, it is relatively easier to specify $(p^*_1,\beta_1)$ based on historical data, and $\beta_2$ based on what is considered a clinically meaningful change in the marginal mean. The specification of $p^*_2$, which is the probability of a latent variable, is difficult due to lack of information on the experimental intervention. With greater interpretability of $q$, it is easier to solicit input from clinical experts. Furthermore, sensitivity analysis that explores a series of potential $q$ values can be communicated back to clinicians as, taking the alcohol-controlling intervention for example, how sample size requirement varies with respect to the relative effect of the intervention on reducing a subject's alcohol consumption versus transforming him/her into an abstiner. 

Finally, given $p_1^*$, if $\beta_2>0$, a larger $q$ is associated with a smaller $p_2^*$, and in turn a smaller variance $\Var(y_{ij})$ under intervention ($r_i=1$) according to (\ref{eqn_var_y}). The association is in the opposite direction if $\beta_2<0$. 

%Finally, although lack of rigorous proof, we observe that the sample size monotonically increase (decrease) with respect to $q \in [0,1]$ if $\beta_2 < 0$ $(\beta_2 >0)$. Therefore, an informative sample size range can be obtained by just considering the cases with $q=0$ and $q=1$.
\subsection{Estimating auxiliary parameters}
The derivation of $N^{(t)}$ and $N^{(z)}$ in Section 2.1 assumes ($p^*_1, p^*_2$) to be known. In actual data analysis they are most likely unknown and need to be estimated. Conventionally, researchers have modeled $p_{ij}$ by a logit model, 
\begin{equation}\label{zero-degenerated}
\log\left(\frac{p_{ij}}{1-p_{ij}}\right) = \alpha_1 + \alpha_2r_i. 
\end{equation}
Statistically speaking, the parameterization by $\bfalpha=(\alpha_1, \alpha_2)'$ is equivalent to that by ($p^*_1, p^*_2$), with $p^*_1=\exp(\alpha_1)/[1+\exp(\alpha_1)]$ and $p^*_2=\exp(\alpha_1+\alpha_2)/[1+\exp(\alpha_1+\alpha_2)]$. Modeling approaches such as (\ref{zero-degenerated}), however, offer greater flexibility to account for additional covariates. In the following we describe how to obtain $(\hat{p}^*_1, \hat{p}^*_2)$ through the  estimation of $\bfalpha$ using the expectation-solution algorithm \citep{kong2015gee}. 

First note that, if $s_{ij}$ ($i = 1,...,N; j = 1,...,m_i$) were observed, parameters $\bfalpha$ could be estimated by solving a GEE equation:
\begin{equation}\label{eqn_GEE_zero_degenerate1}
N^{-\frac{1}{2}}\sum_{i=1}^{N} \left[\frac{\partial
	\boldsymbol{\bfp_i}(\bfalpha)}{\partial \bfalpha}\right]'\bfU_i^{-1}[\bfs_i -
\boldsymbol{\bfp_i}(\bfalpha)] = \bf0.
\end{equation}
Here $\boldsymbol{\bfp_i}(\bfalpha) = p_i\mathbbm{1}_{m_i}$, $\bfs_i =
(s_{i1},\cdots,s_{im_i})'$,
and $\bfU_i$ is a $m_i \times m_i$ diagonal matrix with all
diagonal elements being $p_i(1-p_i)$. 

Since $s_{ij}$ is not observed when $y_{ij} = 0$, the solutions offered by (\ref{eqn_GEE_zero_degenerate1}) is not directly
applicable. Instead, the expectation-solution algorithm can be employed which replaces $s_{ij}$ in (\ref{eqn_GEE_zero_degenerate1}) with $d_{ij}$, its conditional mean given $\{\bfy, \bfbeta, \bfalpha\}$. Specifically,
\begin{eqnarray}\label{condition_mean}
d_{ij} &=& \mbox{Prob}(s_{ij} = 1|\bfy, \bfbeta, \bfalpha),\\\nonumber
&=& \left\{\frac{\mbox{Prob}(s_{ij} = 1,y_{ij}=0|\bfy, \bfbeta,
	\bfalpha)}{\mbox{Prob}(y_{ij} = 0|\bfy, \bfbeta, \bfalpha)}\right\}I_{\{y_{ij}=0\}},\\\nonumber
&=&
\left\{1+ \frac{(1-p_{ij})\exp(-\lambda_{ij})}{p_{ij}}\right\}^{-1}I_{\{y_{ij}=0\}},\\\nonumber
&=&\frac{p_i}{p_i+(1-p_i)\exp(\lambda_i)}I{\{y_{ij}=0\}}. 
\end{eqnarray}
Defining $\bfd_i=(d_{i1}, \cdots, d_{im_i})'$, we modify (\ref{eqn_GEE_zero_degenerate1}) to
\begin{equation}\label{eqn_GEE_zero_degenerate2}
N^{-\frac{1}{2}}\sum_{i=1}^{N} \left[\frac{\partial
	\boldsymbol{\bfp_i}(\bfalpha)}{\partial \bfalpha}\right]'\bfU_i^{-1}[\bfd_i -
\boldsymbol{\bfp_i}(\bfalpha)] = \bf0.
\end{equation} 
Finally, the complete estimation procedure is: 
\begin{enumerate}
	\item Obtain $\hat{\bfalpha}^{(0)}$, the initial value, by running a logistic regression using $\{I_{\{y_{ij} = 0\}}\}$ as the response variable. 
	\item Plug $\hat{\bfalpha}^{(0)}$ into (\ref{eqn_GEE}) to obtain $\hat{\bfbeta}^{(1)}$. Given
	$\{\hat{\bfbeta}^{(1)}, \hat{\bfalpha}^{(0)} \}$, calculate $d_{ij}^{(1)}$'s using (\ref{condition_mean}). Then plug $d_{ij}^{(1)}$'s into (\ref{eqn_GEE_zero_degenerate2}) to obtain $\hat{\bfalpha}^{(1)}$. 
	\item Repeat Step 2 until the estimators converge.      
\end{enumerate}
Then $\hat{p}^*_1=\exp(\hat{\alpha}_1)/[1+\exp(\hat{\alpha}_1)]$ and $\hat{p}^*_2=\exp(\hat{\alpha}_1+\hat{\alpha}_2)/[1+\exp(\hat{\alpha}_1+\hat{\alpha}_2)]$. \cite{kong2015gee} also showed that the expectation-solution algorithm can be employed to estimate $\rho_s$ and $\rho_u$.

\subsection{Addressing the issue of small sample sizes}

In CRTs, the number of clusters ($N$) is often limited \citep{ivers2011impact}. In such cases the sandwich-type variance estimator $\bfSigma_{N}$
is known to be biased downwards, leading to an inflated type I error
\citep{li2015small}.
Alternatively, $\sigma_2^2$ can be estimated using re-sampling based
methods, such as the Jackknife approach \citep{efron1994introduction}. Many researchers have shown that better inference results can be obtained using re-sampling methods \citep{sherman1997comparison,hussey2007design}. 
Let $\bfSigma_{N}^{(Jack)}$ denote the estimate of $ \bfSigma$ using the
Jackknife approach. It is calculated by 
\begin{equation}\label{Jackknife}
\bfSigma_{N}^{(Jack)} = \frac{N-2}{N}\sum_{i=1}^N
(\hat{\bfbeta}^{(-i)}-\hat{\bfbeta})(\hat{\bfbeta}^{(-i)}-\hat{\bfbeta})',
\end{equation}
where $\hat{\bfbeta}^{(-i)}$ denotes the estimate of $\bfbeta$ based on data
excluding the $i$th cluster. Importantly, we perform the re-sampling step at cluster level instead of patient level, so that within-cluster correlation is preserved \citep{sherman1997comparison}. 
We denote the sandwich-type variance estimation approach as
``GEE-Naive'', and the Jackknife approach by (\ref{Jackknife}) as
``GEE-Jackknife''. 
\section{Simulation}
We conduct simulations to assess performance of the proposed sample size method in terms of empirical power and type I error. Suppose
$N$ clusters are randomized 1:1 to the control and intervention arms ($\bar{r}=0.5$). We assume cluster sizes ($m_i$) to be randomly varying, and three distributions are considered: a truncated Poisson distribution with a mean parameter $45$ over a range of $[20,70]$, denoted by TrunPoisson(20,70) with mean $\eta_m \approx 45$ and variance $\sigma_m^2 \approx 44.8$; a discrete uniform distribution (DU) over a range of $[34, 56]$, denoted by DU(34,56) with $\eta_m=45$ and $\sigma_m^2 = 44$; a DU(10,80) distribution with the same mean $\eta_m=45$ but greater variability $\sigma_m^2 = 420$. Define $\bfrho = (\rho_s,\rho_u)$, and two sets of ICCs are explored: $\bfrho =(0.03, 0.03), (0.05, 0.05)$. ICCs of similar magnitude have been frequently reported in CRTs \citep{murray1998design}.
The nominal levels of two-sided type I error and power are set at $\alpha=0.05$ and $1-\gamma$ = 0.8, respectively. We set the regression parameters
$\alpha_1=\beta_1=0$, implying that for the control group, $y_{ij}$ has a $50\%$ ($p^*_1=0.5$) chance of being a structural zero, and the overall mean is $\mu_{ij}=1$%(the corresponding $\lambda_{ij}=2$)
. The goal of the CRT is to assess whether the intervention reduces a count outcome (e.g., the number of falls in nursing homes). We set $\beta_2 = -0.431$, which corresponds to a 0.35 decrease in the overall mean ($\mu_{ij}$ from 1 to 0.65). Finally, five values of $q$ are explored: $q = 0.3, 0.4, 0.5, 0.6$ and $ 0.7$, allowing a sensitivity analysis on the proportion of treatment effect due to change in the probability of structural zeros between groups. Given a particular combination of design configurations, the simulation scheme is described as follows:
\begin{enumerate}
  \item Given parameters $(p^*_1,\beta_2,q)$, compute the value of $p^*_2$.  
	\item Plug the design parameters into (\ref{ss_eqn1}) to compute
	sample size $N^{(z)}$ or further plug $N^{(z)}$ into
	(\ref{ss_eqn3}) to obtain $N^{(t)}$. 
	\item For each scenario, we run $L=2000$ iterations. In the $l$th iteration,
	\begin{enumerate}
		\item Generate a random dataset of $N^{(z)}$ or $N^{(t)}$ clusters under the
		alternative
		hypothesis ($\beta_2=\beta_{20}$). 
		\begin{enumerate}
			\item For each cluster, first generate
			cluster size $m_i$ from the assumed TrunPoisson or
			DU distribution. 
			\item Randomize this cluster to control or intervention. Depending
			on which arm the cluster belongs to, the overall mean $\mu_i$ and the
			probability of being a structural zero $p_i$ are determined. Given
			$\mu_i$ and $p_i$, the mean of the Poisson part is determined (i.e.
			$\lambda_{i} = \frac{\mu_i}{1-p_i}$). 
			\item An $m_i$-length vector of correlated binary variables
			$(s_{i1},...,s_{im_i})$ is generated using the method of
			\cite{qaqish2003family} with marginal probability $p_i$ and ICC
			$\rho_s$. 
			\item An $m_i$-length vector of correlated Poisson variables
			$(u_{i1},...,u_{im_i})$ is obtained by generating $u_{ij} = v_{ij}+v_{i}^{*}$,
			where $ v_{ij}'s$ are random variables from a Poisson distribution with mean $\lambda_i(1-\rho_u)$ and $v_{i}^{*}$ is generated from a Poisson
			distribution with mean $\lambda_i\rho_u$ \citep{mardia1970families}.    
			\item Finally, the $m_i$-length response vector $\bfy_i
			=(y_{i1},...,y_{im_i})'$ is obtained through operation $y_{ij} =
			(1-s_{ij})u_{ij}$ for $j = 1,...,m_i$.   
		\end{enumerate}
		\item Based on the generated dataset, we obtain $\hat{\beta_2}$,
		$\hat{\sigma}_2^{2(Naive)}$, $\hat{\sigma}_2^{2(Jack)}$, respectively. 
	\end{enumerate}
	\item Empirical power of the ``GEE-Naive'' approach is computed as the
	proportion
	of iterations where $|\sqrt{N}
	\frac{\hat{\beta_2}}{\hat{\sigma}_2^{(Naive)}}| > z_{1-0.05/2}$ for $N^{(z)}$
	and $|\sqrt{N}
	\frac{\hat{\beta_2}}{\hat{\sigma}_2^{(Naive)}}| > t_{N^{(t)}-4, 1-0.05/2}$ for
	$N^{(t)}$. The empirical
	powers of the ``GEE-Jackknife'' approach are computed using $\hat{\sigma}_2^{2(Jack)}$.   
	\item Empirical type I error is obtained similarly except for setting $\beta_2=0$ in Step 3(a).
\end{enumerate}
Tables 1-2 present the numbers of clusters $N^{(z)}$ and $N^{(t)}$, empirical type I error, and
empirical power for the ``GEE-Naive'' and ``GEE-Jackknife'' approaches under different combinations of design parameters. Across all scenarios, the numbers of clusters $N^{(z)}$ and $N^{(t)}$ range from $18$ to $30$ and $21$ to $32$, respectively, with $N^{(t)}$ being slightly larger than $N^{(z)}$ in every setting. When the ``GEE-Naive'' approach is paired with $N^{(z)}$, the empirical type I error and power tend to be seriously inflated.  Pairing the ``GEE-Naive'' approach with $N^{(t)}$, or the ``GEE-Jackknife" approach with $N^{(z)}$, leads to slightly better performance, but moderate inflation in type I error persists. Finally, the combination of the ``GEE-Jackknife'' approach and $N^{(t)}$ achieves the best performance, with both type I error and power controlled at nominal levels across all scenarios. Therefore, in practice we recommend calculating sample size using the $t$-distribution-based formula and perform data analysis using the ``GEE-Jackknife'' approach.

Tables 1-2 show a monotone relationship between $q$ and sample size, which is consistent with the theoretical property that if $\beta_2<0$, a larger $q$ is associated with a larger variance $\Var(y_{ij})$ under intervention. We also observe that the proposed sample size is quite robust to change in $q$. In particular, for the sample sizes based on $t$-distribution approximation (i.e., Table 2), the sample sizes are mostly within 5\% from the median (obtained at $q=0.5$) as $q$ varies from $0.3$ to $0.7$. The variation in sample size is smaller under weaker ICCs, i.e., $(\rho_s,\rho_u)=(0.03, 0.03)$. The above observation suggests that in practice, when there is limited prior knowledge, using $q=0.5$ as a default specification might provide a reasonable initial sample size assessment. 
 
The proposed sample size method accommodates random varying cluster sizes through the mean and variance parameters ($\eta_m$ and $\sigma_m^2$). In Tables 1 and 2, the comparison of results between TrunPoisson (20, 70) and  DU(34, 56) represents a sensitivity analysis on cluster size distributions, where the means and variances are comparable but the distributions are different. No significant difference is observed in simulation results between the distributions, suggesting that the proposed sample size method is robust to randomly varying cluster sizes of different distributions. On the other hand, the comparison between DU(34, 56) and DU(10, 80) represents another sensitivity analysis where the cluster sizes follow the same type of distribution (DU) with a common center ($\eta_m=45$), but the variances are different. The results show that larger variability in cluster size leads to larger sample size requirement. 

To demonstrate the consequence of misusing Poisson-based power analysis methods for CRTs with ZIP outcomes, we compare the number of clusters calculated between the proposed method and the Poisson-based approach by \cite{wang2018}, which also accounts for random variability in cluster size. Since \cite{wang2018} only considered normal distribution, we use Equation (\ref{ss_eqn2}), i.e., $N^{(z)}$, to calculate the number of clusters for a fair comparison. %To make the settings comparable, 
For each configuration, we set the Poisson mean equal to the marginal mean of the ZIP model. Recall that ICCs are specified on both the Poisson part and structural zero part for the ZIP model. To obtain a comparable ICC for the Poisson model, we generate a ZIP data set with $10000$ clusters, and then estimate ICC by fitting a Poisson model with an ``exchangeable" correlation structure using the GEE approach. We denote this estimated ICC as $\hat{\rho}^{(Poisson)}$. The resulting sample sizes are presented in Table 3. Mistakenly applying a Poisson-based method to a CRT with a zero-inflated outcome would lead to a severely under-powered clinical trial. 

\section{Application}
We apply the proposed sample size method to a CRT that evaluated the effectiveness of knowledge translation strategies for care team members in long-term care (LTC) settings \citep{kennedy2015successful}. The outcome of interest was the number of falls for senior residents which exhibited zero-inflation. Forty LTC homes were included as clusters and the average number of participants per cluster was 137, with a range of [43, 375]. The study reported that, during a three month follow-up, the average number of falls and the proportion of zeros were $(1.21, 37.2\%)$ in the control arm.

Suppose we want to design a new CRT to investigate the effectiveness of a new intervention on reducing falls in LTC homes, where the power and two-sided type I error are set at 80\% and 5\%, respectively. The design parameters to estimate the required number of clusters $(N)$ are described as follows: We assume the average number of falls and the proportion of zeros to be $(1.21, 37.2\%)$ for the control group, as observed in the original study. It implies that  $p_1^*=12.1\%$ and $\beta_1=0.19$.
We assume that the intervention reduces the average number of falls to 1.01, which gives $\beta_2=-0.18$. 
We consider $q=0.3, 0.4, 0.5, 0.6$ and $0.7$ for  sensitivity analysis, corresponding to $p_2^*=20.0\%, 22.5\%, 24.9\%, 27.2\%$ and $29.4\%$, respectively.
We further assume ICCs: $\rho_s=\rho_u=0.05$. Suppose the variability in cluster sizes is relatively small, say $m_i \sim \text{DU}[127, 147]$, the required number of clusters is 53 for $q=0.03$, 54 for $q$ between $0.4$ and $0.6$, and 55 for $q=0.7$, respectively, based on the $t$-distribution approximation (\ref{ss_eqn3}). For a larger variability like the original study, say $m_i \sim \text{DU}[37, 237]$, the required number of clusters increases to 61 for $q=0.3$ and $0.4$, and 62 otherwise.

\section{Conclusion}
In this study we present a sample size method for CRTs with zero-inflated count outcomes. It is developed based on GEE regression directly modeling the marginal mean of a ZIP outcome, which avoids the challenge of testing two intervention effects under traditional modeling approaches. We derive closed-form sample size formulas that properly account for zero inflation, ICCs due to clustering, unbalanced randomization, and variability in cluster size. We also introduce a new parameter $q$, which provides a straightforward decomposition of the overall intervention effect to facilitate communication with clinicians, as well as a natural framework to conduct sensitivity analysis.

 %\textcolor{red}{Although we only include the
%intervention indicator in the model (i.e.,  Equations (1) and (2)) for sample size calculation, the algorithm is general in the sense that it is straightforward to incorporate additional covariates (repeated?). 

%In practice, it is common that the number of clusters is relatively small \citep{ivers2011impact} for CRTs. To address the concerns that standard asymptotic properties for GEE may be poor under small sample sizes, we implement two adaptions. First, we relax dependence on the asymptotic normal approximation in the sample size derivation by assuming a more robust \textit{t}-distribution, which has demonstrated good performance for other types of CRT. Second, \textcolor{red}{in the data analysis stage, }it is known that the GEE sandwich-type estimator may underestimate variances for CRTs with small sample sizes, and we provide an alternative variance estimator based on the Jackknife resampling approach. 
We evaluate the performance of the proposed sample size method through extensive simulation. The results show that the combination of calculating sample size based on the \textit{t}-distribution ($N^{(t)}$) in experimental design, and implementing the ``GEE-Jackknife”  approach in data analysis, adequately controls the type I error and power at their nominal levels across all scenarios considered. We further show that traditional power analysis methods based on Poisson distribution tend to seriously underestimate sample sizes when the CRT has a zero-inflated count outcome.

The proposed sample size method is developed under the ZIP framework. In future research we will investigate its extension to other zero-inflated count models, such as the zero-inflated negative binomial model and the hurdle model \citep{mullahy1986specification}. 

\bibliography{clusterCount}

@article{liang1986,
	title={Longitudinal data analysis using generalized linear models},
	author={Liang, Kung-Yee and Zeger, Scott L},
	journal={Biometrika},
	volume={73},
	number={1},
	pages={13--22},
	year={1986},
	publisher={Oxford University Press}
}

@article{amatya2013,
	title={Sample size determination for clustered count data},
	author={Amatya, Anup and Bhaumik, Dulal and Gibbons, Robert D},
	journal={Statistics in Medicine},
	volume={32},
	number={24},
	pages={4162--4179},
	year={2013},
	publisher={Wiley Online Library}
}

@article{wang2018,
	title={Sample size calculation for count outcomes in cluster randomization trials with varying cluster sizes},
	author={Wang, Jijia and Zhang, Song and Ahn, Chul},
	journal={Communications in Statistics-Theory and Methods},
	pages={1--9},
	year={2018},
	publisher={Taylor \& Francis}
}

@article{ivers2011impact,
	title={Impact of CONSORT extension for cluster randomised trials on quality of reporting and study methodology: review of random sample of 300 trials, 2000-8},
	author={Ivers, NM and Taljaard, M and Dixon, Stephanie and Bennett, C and McRae, A and Taleban, J and Skea, Z and Brehaut, JC and Boruch, RF and Eccles, MP and others},
	journal={BMJ},
	volume={343},
	pages={d5886},
	year={2011},
	publisher={British Medical Journal Publishing Group}
}

@article{li2015small,
	title={Small sample performance of bias-corrected sandwich estimators for cluster-randomized trials with binary outcomes},
	author={Li, Peng and Redden, David T},
	journal={Statistics in Medicine},
	volume={34},
	number={2},
	pages={281--296},
	year={2015},
	publisher={Wiley Online Library}
}

@book{murray1998design,
	title={Design and analysis of group-randomized trials},
	author={Murray, David M},
	volume={29},
	year={1998},
	publisher={Monographs in Epidemiology \& B}
}

@book{eldridge2012practical,
	title={A practical guide to cluster randomised trials in health services research},
	author={Eldridge, Sandra and Kerry, Sally},
	volume={120},
	year={2012},
	publisher={John Wiley \& Sons}
}

@article{roig2010cluster,
	title={Cluster randomized trial in smoking cessation with intensive advice in diabetic patients in primary care. ITADI Study},
	author={Roig, Lydia and Perez, Santiago and Prieto, Gemma and Martin, Carlos and Advani, Mamta and Armengol, Angelina and Roura, Pilar and Manresa, Josep Maria and Briones, Elena},
	journal={BMC public health},
	volume={10},
	number={1},
	pages={58},
	year={2010},
	publisher={BioMed Central}
}

@article{fairall2005effect,
	title={Effect of educational outreach to nurses on tuberculosis case detection and primary care of respiratory illness: pragmatic cluster randomised controlled trial},
	author={Fairall, Lara R and Zwarenstein, Merrick and Bateman, Eric D and Bachmann, Max and Lombard, Carl and Majara, Bosielo P and Joubert, Gina and English, Rene G and Bheekie, Angeni and van Rensburg, Dingie and others},
	journal={Bmj},
	volume={331},
	number={7519},
	pages={750--754},
	year={2005},
	publisher={British Medical Journal Publishing Group}
}

@book{ahn2014sample,
	title={Sample size calculations for clustered and longitudinal outcomes in clinical research},
	author={Ahn, Chul and Heo, Moonseoung and Zhang, Song},
	year={2014},
	publisher={Chapman and Hall/CRC}
}

@book{efron1994introduction,
	title={An Introduction to the Bootstrap},
	author={Efron, Bradley and Tibshirani, Robert J},
	year={1994},
	publisher={CRC press}
}

@article{sherman1997comparison,
	title={A comparison between bootstrap methods and generalized estimating equations for correlated outcomes in generalized linear models},
	author={Sherman, Michael and Cessie, Saskia le},
	journal={Communications in Statistics-Simulation and Computation},
	volume={26},
	number={3},
	pages={901--925},
	year={1997},
	publisher={Taylor \& Francis}
}

@article{hussey2007design,
	title={Design and analysis of stepped wedge cluster randomized trials},
	author={Hussey, Michael A and Hughes, James P},
	journal={Contemporary Clinical Trials},
	volume={28},
	number={2},
	pages={182--191},
	year={2007},
	publisher={Elsevier}
}

@article{liu2018relative,
	title={Relative efficiency of unequal versus equal cluster sizes in cluster randomized trials using generalized estimating equation models},
	author={Liu, Jingxia and Colditz, Graham A},
	journal={Biometrical Journal},
	volume={60},
	number={3},
	pages={616--638},
	year={2018},
	publisher={Wiley Online Library}
}

@article{murray2004design,
	title={Design and analysis of group-randomized trials: a review of recent methodological developments},
	author={Murray, David M and Varnell, Sherri P and Blitstein, Jonathan L},
	journal={American Journal of Public Health},
	volume={94},
	number={3},
	pages={423--432},
	year={2004},
	publisher={American Public Health Association}
}

@article{qaqish2003family,
	title={A family of multivariate binary distributions for simulating correlated binary variables with specified marginal means and correlations},
	author={Qaqish, Bahjat F},
	journal={Biometrika},
	volume={90},
	number={2},
	pages={455--463},
	year={2003},
	publisher={Oxford University Press}
}

@article{kong2015gee,
	title={GEE type inference for clustered zero-inflated negative binomial regression with application to dental caries},
	author={Kong, Maiying and Xu, Sheng and Levy, Steven M and Datta, Somnath},
	journal={Computational statistics \& data analysis},
	volume={85},
	pages={54--66},
	year={2015},
	publisher={Elsevier}
}

@article{tang2017closed,
	title={Closed-form REML estimators and sample size determination for mixed effects models for repeated measures under monotone missingness},
	author={Tang, Yongqiang},
	journal={Statistics in medicine},
	volume={36},
	number={13},
	pages={2135--2147},
	year={2017},
	publisher={Wiley Online Library}
}

@book{mardia1970families,
	title={Families of bivariate distributions},
	author={Mardia, Kantilal Varichand},
	volume={27},
	year={1970},
	publisher={Lubrecht \& Cramer Ltd}
}

@article{martin2004multicentre,
  title={Multicentre, cluster-randomized clinical trial of algorithms for critical-care enteral and parenteral therapy (ACCEPT)},
  author={Martin, Claudio M and Doig, Gordon S and Heyland, Daren K and Morrison, Teresa and Sibbald, William J and others},
  journal={Cmaj},
  volume={170},
  number={2},
  pages={197--204},
  year={2004},
  publisher={Can Med Assoc}
}

@article{kennedy2015successful,
  title={Successful knowledge translation intervention in long-term care: final results from the vitamin D and osteoporosis study (ViD OS) pilot cluster randomized controlled trial},
  author={Kennedy, Courtney C and Ioannidis, George and Thabane, Lehana and Adachi, Jonathan D and Marr, Sharon and Giangregorio, Lora M and Morin, Suzanne N and Crilly, Richard G and Josse, Robert G and Lohfeld, Lynne and others},
  journal={Trials},
  volume={16},
  number={1},
  pages={214},
  year={2015},
  publisher={BioMed Central}
}

@article{yang2009testing,
  title={Testing overdispersion in the zero-inflated Poisson model},
  author={Yang, Zhao and Hardin, James W and Addy, Cheryl L},
  journal={Journal of Statistical Planning and Inference},
  volume={139},
  number={9},
  pages={3340--3353},
  year={2009},
  publisher={Elsevier}
}

@article{mullahy1986specification,
  title={Specification and testing of some modified count data models},
  author={Mullahy, John},
  journal={Journal of econometrics},
  volume={33},
  number={3},
  pages={341--365},
  year={1986},
  publisher={Elsevier}
}

@article{preisser2012review,
  title={Review and recommendations for zero-inflated count regression modeling of dental caries indices in epidemiological studies},
  author={Preisser, John S and Stamm, John W and Long, D Leann and Kincade, Megan E},
  journal={Caries research},
  volume={46},
  number={4},
  pages={413--423},
  year={2012},
  publisher={Karger Publishers}
}

@article{famoye2006zero,
  title={Zero-inflated generalized Poisson regression model with an application to domestic violence data},
  author={Famoye, Felix and Singh, Karan P},
  journal={Journal of Data Science},
  volume={4},
  number={1},
  pages={117--130},
  year={2006}
}

@article{long2014marginalized,
  title={A marginalized zero-inflated Poisson regression model with overall exposure effects},
  author={Long, D Leann and Preisser, John S and Herring, Amy H and Golin, Carol E},
  journal={Statistics in medicine},
  volume={33},
  number={29},
  pages={5151--5165},
  year={2014},
  publisher={Wiley Online Library}
}

@book{cameron2013regression,
  title={Regression analysis of count data},
  author={Cameron, A Colin and Trivedi, Pravin K},
  volume={53},
  year={2013},
  publisher={Cambridge university press}
}

@book{agresti2003categorical,
  title={Categorical data analysis},
  author={Agresti, Alan},
  volume={482},
  year={2003},
  publisher={John Wiley \& Sons}
}

@article{lewsey2004utility,
  title={The utility of the zero-inflated Poisson and zero-inflated negative binomial models: a case study of cross-sectional and longitudinal DMF data examining the effect of socio-economic status},
  author={Lewsey, James D and Thomson, William M},
  journal={Community dentistry and oral epidemiology},
  volume={32},
  number={3},
  pages={183--189},
  year={2004},
  publisher={Wiley Online Library}
}

@article{lambert1992zero,
  title={Zero-inflated Poisson regression, with an application to defects in manufacturing},
  author={Lambert, Diane},
  journal={Technometrics},
  volume={34},
  number={1},
  pages={1--14},
  year={1992},
  publisher={Taylor \& Francis}
}

@article{moghimbeigi2008multilevel,
  title={Multilevel zero-inflated negative binomial regression modeling for over-dispersed count data with extra zeros},
  author={Moghimbeigi, Abbas and Eshraghian, Mohammed Reza and Mohammad, Kazem and Mcardle, Brian},
  journal={Journal of Applied Statistics},
  volume={35},
  number={10},
  pages={1193--1202},
  year={2008},
  publisher={Taylor \& Francis}
}

@article{li2019sample,
  title={Sample size calculation for clinical trials with correlated count measurements based on the negative binomial distribution},
  author={Li, Dateng and Zhang, Song and Cao, Jing},
  journal={Statistics in Medicine},
  volume={38},
  number={28},
  pages={5413--5427},
  year={2019},
  publisher={Wiley Online Library}
}

@article{beckett2014zero,
  title={Zero-inflated Poisson (ZIP) distribution: parameter estimation and applications to model data from natural calamities},
  author={Beckett, Sadie and Jee, Joshua and Ncube, Thapelo and Pompilus, Sophia and Washington, Quintel and Singh, Anshuman and Pal, Nabendu},
  journal={Involve, a Journal of Mathematics},
  volume={7},
  number={6},
  pages={751--767},
  year={2014},
  publisher={Mathematical Sciences Publishers}
}
%\bibliography{/Users/yszhang/Song/settings/latex/myref}
%\bibliographystyle{/Users/yszhang/Song/settings/latex/agsm}

%\bibliography{/Volumes/SHARED/Song/settings/latex/myref}
%\bibliography{/Users/yszhang/lw/song/settings/latex/myref}
%\bibliographystyle{/Users/yszhang/lw/song/settings/latex/agsm}
%\bibliography{/home/yszhang/lw/Song/settings/latex/myref}
%\bibliographystyle{/home/yszhang/lw/Song/settings/latex/agsm}
%\bibliography{myref}
\bibliographystyle{agsm}
\section*{Appendix A. Derivation of $\bfV_1$ and $\bfV_2$}
$\bfV_1$ and $\bfV_2$ are derived in the similar manner. Here we present the derivation of $\bfV_1$. First rewrite $\bfV_1$ as
\begin{eqnarray}
\bfV_1 &=&  \left(\begin{array}{cc}
1 & 0\\\nonumber
0 & 0
\end{array}\right)\frac{1}{\left(1+\frac{p^*_1}{1-p^*_1}\mu^*_1	\right)^2}\sum_{m \in
	\mathcal{M}}g(m)
E\left[\sum_{j=1}^m\sum_{j'=1}^m(y_{ij}-\mu^*_{1})(y_{ij'}-\mu^*_{1})\right].
\end{eqnarray}
For $E\left[\sum_{j=1}^m\sum_{j'=1}^m(y_{ij}-\mu^*_{1})(y_{ij'}-\mu^*_{1})\right]$,
we have 
\begin{equation*}
E\left[\sum_{j=1}^m\sum_{j'=1}^m(y_{ij}-\mu^*_{1})(y_{ij'}-\mu^*_{1})\right] =
\sum_{j=1}^{m}E(y_{ij}-\mu^*_{1})^2+2\sum_{j=1}^{m-1}\sum_{j'=j+1}^{m}E\left[(y_{ij}-\mu^*_{1})(y_{ij'}-\mu^*_{1})\right].
\end{equation*} 
It is clear that 
\begin{equation*}
\sum_{j=1}^{m}E(y_{ij}-\mu^*_{1})^2 =\sum_{j=1}^{m} \Var(y_{ij}) =
m\left[\mu^*_{1}+\frac{p^*_1}{1-p^*_1}\mu_{1}^{*2}\right].
\end{equation*}
On the other hand, we have
\begin{equation*}
2\sum_{j=1}^{m-1}\sum_{j'=j+1}^{m}E\left[(y_{ij}-\mu^*_{1})(y_{ij'}-\mu^*_{1})\right]=2\sum_{j=1}^{m-1}\sum_{j'=j+1}^{m}\Cov(y_{ij},y_{ij'})=2(m^2-m)\Cov(y_{ij},y_{ij'}).
\end{equation*} 
If $s_{ij} =
s_{ij'}=1$, 
\begin{equation*}
E\left[(y_{ij}-\mu^*_{1})(y_{ij'}-\mu^*_{1})\right]
=E\left[(0-\mu^*_{1})(0-\mu^*_{1})\right]= \mu_1^{*2};
\end{equation*}
if $s_{ij} = 1, s_{ij'}=0$, 
\begin{eqnarray*}
	E\left[(y_{ij}-\mu^*_{1})(y_{ij'}-\mu^*_{1})\right] &=&
	E\left[(0-\mu^*_{1})(u_{ij'}-\mu^*_{1})\right] \\\nonumber
	&=&-\mu^*_1E\left[u_{ij'}-\lambda^*_1+p_1\lambda^*_1\right]\\\nonumber
	&=& -\mu_1^{*2}\frac{p^*_1}{1-p^*_1},\nonumber
\end{eqnarray*}
where $\lambda^*_1 = \frac{\mu^*_1}{1-p^*_1}$;\\
if $s_{ij} = s_{ij'}=0$, 
\begin{eqnarray*}
	E\left[(y_{ij}-\mu^*_{1})(y_{ij'}-\mu^*_{1})\right] &=&
	E\left[(u_{ij}-\mu^*_{1})(u_{ij'}-\mu^*_{1})\right] \\\nonumber
	&=&E\left[(u_{ij}-\lambda^*_1+p^*_1\lambda^*_1)(u_{ij'}-\lambda^*_1+p^*_1\lambda^*_1)\right]\\\nonumber
	&=& \rho_u\lambda^*_1 + p_1^{*2}\lambda_1^{*2} \\\nonumber
	&=& \frac{\rho_u\mu^*_1}{1-p^*_1}+\frac{p_1^{*2}\mu_1^{*2}}{(1-p^*_1)^2}.\nonumber
\end{eqnarray*}
It is easy to verify that
$\mbox{Prob}(s_{ij}=s_{ij'}=1)=p_1^{*2}+p^*_1(1-p^*_1)\rho_s$; $\mbox{Prob}(s_{ij}=0,
s_{ij'}=1)=(1-p^*_1)p^*_1(1-\rho_s)$;
$\mbox{Prob}(s_{ij}=s_{ij'}=0)=(1-p^*_1)(1-p^*_1+\rho_sp^*_1)$. 

In addition, it is clear that $\sum_{m \in \mathcal{M}}g(m)m=\eta_m$ and $\sum_{m \in \mathcal{M}}g(m)(m^2-m)= \eta_m^2+\sigma_m^2-\eta_m$.
Putting together, we have  
\begin{eqnarray}
\bfV_1 &=&  \left(\begin{array}{cc}
1 & 0\\\nonumber
0 & 0
\end{array}\right)\left[\eta_m\frac{\mu^*_1+\frac{p^*_1}{1-p^*_1}\mu_1^{*2}}{(1+\frac{p^*_1}{1-p^*_1}\mu^*_1)^2}
+ (\eta_m^2+\sigma_m^2-\eta_m)\frac{\zeta_1}{(1+\frac{p^*_1}{1-p^*_1}\mu^*_1)^2}\right],
\end{eqnarray}
with 
\begin{equation*}
\zeta_1 =
\mu^*_1\left[\frac{p^*_1}{1-p^*_1}\rho_s-2(\mu^*_1+2)p_1^{*2}(\rho_s-1)+p^*_1(\rho_s-1)(\rho_u + 2)+\rho_u\right].
\end{equation*}
Hence we complete the derivation of $\bfV_1$. 

\begingroup
\renewcommand{\arraystretch}{0.8}
\begin{table}[]
	\centering
	\caption{Simulation: empirical type I error and power for $N^{(z)}$}
	\vspace{0.3cm}
\begin{tabular}{|cccccccc|}
\hline
                                            &                      &                         &           & \multicolumn{2}{c}{GEE-Naive} & \multicolumn{2}{c|}{GEE-Jackknife} \\
                                            & $(\rho_s, \rho_u)$   & $q$                     & $N^{(z)}$ & Type I Error      & Power     & Type I Error        & Power        \\ \hline
\multicolumn{1}{|c|}{}                      &                      & 0.3                     & 18        & 0.085             & 0.846     & 0.067               & 0.814        \\
\multicolumn{1}{|c|}{}                      &                      & 0.4                     & 19        & 0.079             & 0.858     & 0.062               & 0.818        \\
\multicolumn{1}{|l|}{}                      & (0.03, 0.03)         & \multicolumn{1}{l}{0.5} & 19        & 0.080             & 0.857     & 0.063               & 0.821        \\
\multicolumn{1}{|l|}{}                      & \multicolumn{1}{l}{} & \multicolumn{1}{l}{0.6} & 20        & 0.082             & 0.851     & 0.065               & 0.823        \\
\multicolumn{1}{|c|}{TrunPoisson(45,20,70)} &                      & 0.7                     & 20        & 0.082             & 0.844     & 0.065               & 0.815        \\ \cline{2-8} 
\multicolumn{1}{|c|}{}                      &                      & 0.3                     & 24        & 0.075             & 0.835     & 0.061               & 0.807        \\
\multicolumn{1}{|l|}{}                      & \multicolumn{1}{l}{} & 0.4                     & 25        & 0.071             & 0.849     & 0.061               & 0.823        \\
\multicolumn{1}{|c|}{}                      & (0.05,0.05)          & \multicolumn{1}{l}{0.5} & 25        & 0.071             & 0.843     & 0.061               & 0.818        \\
\multicolumn{1}{|l|}{}                      & \multicolumn{1}{l}{} & \multicolumn{1}{l}{0.6} & 26        & 0.068             & 0.849     & 0.055               & 0.825        \\
\multicolumn{1}{|c|}{}                      &                      & 0.7                     & 27        & 0.068             & 0.856     & 0.054               & 0.827        \\ \hline
\multicolumn{1}{|c|}{}                      &                      & 0.3                     & 18        & 0.081             & 0.848     & 0.063               & 0.814        \\
\multicolumn{1}{|l|}{}                      & \multicolumn{1}{l}{} & 0.4                     & 19        & 0.083             & 0.844     & 0.065               & 0.817        \\
\multicolumn{1}{|c|}{}                      & (0.03, 0.03)         & \multicolumn{1}{l}{0.5} & 19        & 0.083             & 0.845     & 0.065               & 0.805        \\
\multicolumn{1}{|l|}{}                      & \multicolumn{1}{l}{} & \multicolumn{1}{l}{0.6} & 20        & 0.086             & 0.849     & 0.066               & 0.823        \\
\multicolumn{1}{|c|}{DU(34, 56)}            &                      & 0.7                     & 20        & 0.086             & 0.848     & 0.066               & 0.817        \\ \cline{2-8} 
\multicolumn{1}{|c|}{}                      &                      & 0.3                     & 24        & 0.077             & 0.836     & 0.061               & 0.805        \\
\multicolumn{1}{|l|}{}                      & \multicolumn{1}{l}{} & 0.4                     & 25        & 0.075             & 0.852     & 0.063               & 0.829        \\
\multicolumn{1}{|c|}{}                      & (0.05,0.05)          & \multicolumn{1}{l}{0.5} & 25        & 0.075             & 0.852     & 0.063               & 0.820        \\
\multicolumn{1}{|l|}{}                      & \multicolumn{1}{l}{} & \multicolumn{1}{l}{0.6} & 26        & 0.074             & 0.843     & 0.062               & 0.809        \\
\multicolumn{1}{|c|}{}                      &                      & 0.7                     & 27        & 0.071             & 0.841     & 0.064               & 0.818        \\ \hline
\multicolumn{1}{|c|}{}                      &                      & 0.3                     & 20        & 0.094             & 0.851     & 0.075               & 0.813        \\
\multicolumn{1}{|l|}{}                      & \multicolumn{1}{l}{} & 0.4                     & 20        & 0.095             & 0.849     & 0.073               & 0.804        \\
\multicolumn{1}{|c|}{}                      & (0.03, 0.03)         & \multicolumn{1}{l}{0.5} & 21        & 0.093             & 0.859     & 0.069               & 0.818        \\
\multicolumn{1}{|l|}{}                      & \multicolumn{1}{l}{} & \multicolumn{1}{l}{0.6} & 21        & 0.093             & 0.851     & 0.069               & 0.803        \\
\multicolumn{1}{|c|}{DU(10, 80)}            &                      & 0.7                     & 22        & 0.086             & 0.847     & 0.069               & 0.804        \\ \cline{2-8} 
\multicolumn{1}{|c|}{}                      &                      & 0.3                     & 27        & 0.074             & 0.828     & 0.051               & 0.794        \\
\multicolumn{1}{|l|}{}                      & \multicolumn{1}{l}{} & 0.4                     & 28        & 0.073             & 0.837     & 0.054               & 0.802        \\
\multicolumn{1}{|c|}{}                      & (0.05,0.05)          & \multicolumn{1}{l}{0.5} & 28        & 0.073             & 0.829     & 0.054               & 0.799        \\
\multicolumn{1}{|l|}{}                      & \multicolumn{1}{l}{} & \multicolumn{1}{l}{0.6} & 29        & 0.067             & 0.846     & 0.054               & 0.811        \\
\multicolumn{1}{|c|}{}                      &                      & 0.7                     & 30        & 0.062             & 0.835     & 0.049               & 0.804        \\ \hline
\end{tabular}
\end{table}
\endgroup

\begingroup
\renewcommand{\arraystretch}{0.8}
\begin{table}[]
	\centering
	\caption{Simulation: empirical type I error and power for $N^{(t)}$}
	\vspace{0.3cm}
\begin{tabular}{|cccccccc|}
\hline
                                            &                      &                         &           & \multicolumn{2}{c}{GEE-Naive} & \multicolumn{2}{c|}{GEE-Jackknife} \\
                                            & $(\rho_s, \rho_u)$   & $q$                     & $N^{(t)}$ & Type I Error      & Power     & Type I Error        & Power        \\ \hline
\multicolumn{1}{|c|}{}                      &                      & 0.3                     & 21        & 0.066             & 0.862     & 0.053               & 0.837        \\
\multicolumn{1}{|c|}{}                      &                      & 0.4                     & 21        & 0.066             & 0.866     & 0.053               & 0.827        \\
\multicolumn{1}{|l|}{}                      & (0.03, 0.03)         & \multicolumn{1}{l}{0.5} & 22        & 0.064             & 0.873     & 0.051               & 0.834        \\
\multicolumn{1}{|l|}{}                      & \multicolumn{1}{l}{} & \multicolumn{1}{l}{0.6} & 22        & 0.064             & 0.851     & 0.051               & 0.816        \\
\multicolumn{1}{|c|}{TrunPoisson(45,20,70)} &                      & 0.7                     & 22        & 0.064             & 0.850     & 0.051               & 0.820        \\ \cline{2-8} 
\multicolumn{1}{|c|}{}                      &                      & 0.3                     & 27        & 0.053             & 0.855     & 0.039               & 0.825        \\
\multicolumn{1}{|l|}{}                      & \multicolumn{1}{l}{} & 0.4                     & 27        & 0.053             & 0.842     & 0.039               & 0.819        \\
\multicolumn{1}{|c|}{}                      & (0.05,0.05)          & \multicolumn{1}{l}{0.5} & 28        & 0.052             & 0.850     & 0.044               & 0.826        \\
\multicolumn{1}{|l|}{}                      & \multicolumn{1}{l}{} & \multicolumn{1}{l}{0.6} & 28        & 0.052             & 0.857     & 0.044               & 0.836        \\
\multicolumn{1}{|c|}{}                      &                      & 0.7                     & 29        & 0.050             & 0.851     & 0.043               & 0.833        \\ \hline
\multicolumn{1}{|c|}{}                      &                      & 0.3                     & 21        & 0.064             & 0.859     & 0.049               & 0.830        \\
\multicolumn{1}{|l|}{}                      & \multicolumn{1}{l}{} & 0.4                     & 21        & 0.064             & 0.852     & 0.049               & 0.822        \\
\multicolumn{1}{|c|}{}                      & (0.03, 0.03)         & \multicolumn{1}{l}{0.5} & 21        & 0.066             & 0.845     & 0.051               & 0.812        \\
\multicolumn{1}{|l|}{}                      & \multicolumn{1}{l}{} & \multicolumn{1}{l}{0.6} & 22        & 0.063             & 0.850     & 0.054               & 0.823        \\
\multicolumn{1}{|c|}{DU(34, 56)}            &                      & 0.7                     & 22        & 0.063             & 0.847     & 0.054               & 0.808        \\ \cline{2-8} 
\multicolumn{1}{|c|}{}                      &                      & 0.3                     & 27        & 0.063             & 0.843     & 0.055               & 0.808        \\
\multicolumn{1}{|l|}{}                      & \multicolumn{1}{l}{} & 0.4                     & 27        & 0.063             & 0.850     & 0.055               & 0.821        \\
\multicolumn{1}{|c|}{}                      & (0.05,0.05)          & \multicolumn{1}{l}{0.5} & 28        & 0.064             & 0.858     & 0.058               & 0.834        \\
\multicolumn{1}{|l|}{}                      & \multicolumn{1}{l}{} & \multicolumn{1}{l}{0.6} & 28        & 0.064             & 0.841     & 0.058               & 0.819        \\
\multicolumn{1}{|c|}{}                      &                      & 0.7                     & 29        & 0.063             & 0.853     & 0.054               & 0.828        \\ \hline
\multicolumn{1}{|c|}{}                      &                      & 0.3                     & 22        & 0.068             & 0.852     & 0.054               & 0.818        \\
\multicolumn{1}{|l|}{}                      & \multicolumn{1}{l}{} & 0.4                     & 23        & 0.061             & 0.860     & 0.041               & 0.819        \\
\multicolumn{1}{|c|}{}                      & (0.03, 0.03)         & \multicolumn{1}{l}{0.5} & 23        & 0.066             & 0.847     & 0.046               & 0.810        \\
\multicolumn{1}{|l|}{}                      & \multicolumn{1}{l}{} & \multicolumn{1}{l}{0.6} & 24        & 0.068             & 0.851     & 0.049               & 0.814        \\
\multicolumn{1}{|c|}{DU(10, 80)}            &                      & 0.7                     & 24        & 0.064             & 0.842     & 0.046               & 0.807        \\ \cline{2-8} 
\multicolumn{1}{|c|}{}                      &                      & 0.3                     & 29        & 0.056             & 0.836     & 0.042               & 0.803        \\
\multicolumn{1}{|l|}{}                      & \multicolumn{1}{l}{} & 0.4                     & 30        & 0.052             & 0.835     & 0.041               & 0.812        \\
\multicolumn{1}{|c|}{}                      & (0.05,0.05)          & \multicolumn{1}{l}{0.5} & 30        & 0.052             & 0.834     & 0.041               & 0.803        \\
\multicolumn{1}{|l|}{}                      & \multicolumn{1}{l}{} & \multicolumn{1}{l}{0.6} & 31        & 0.052             & 0.841     & 0.042               & 0.808        \\
\multicolumn{1}{|c|}{}                      &                      & 0.7                     & 32        & 0.053             & 0.844     & 0.040               & 0.809        \\ \hline
\end{tabular}
\end{table}
\endgroup

\begin{table}[]
	\centering
    	\caption{Comparison of $N^{(ZIP)}$ and $N^{(Poisson)}$ under randomly varying cluster sizes. Here $N^{(ZIP)}$ is the number of clusters calculated from formula (\ref{ss_eqn2}), and $N^{(Poisson)}$ is calculated based on the Poisson distribution \citep{wang2018}.  $\hat{\rho}^{(Poisson)}$ is the estimated ICC for the Poisson distribution.} 
    	
    	\vspace{0.3cm}
\begin{tabular}{|cccccc|}
\hline
                                 & $(\rho_s, \rho_u)$                & $q$                     & $N^{(ZIP)}$ & $\hat{\rho}^{(Poisson)}$ & $N^{(Poisson)}$ \\ \hline
\multicolumn{1}{|c|}{}           & \multicolumn{1}{c|}{}             & 0.3                     & 18          & 0.022                & 10              \\
\multicolumn{1}{|l|}{}           & \multicolumn{1}{l|}{}             & \multicolumn{1}{l}{0.4} & 19          & 0.021                & 10              \\
\multicolumn{1}{|c|}{}           & \multicolumn{1}{c|}{(0.03, 0.03)} & 0.5                     & 19          & 0.023                & 10              \\
\multicolumn{1}{|l|}{}           & \multicolumn{1}{l|}{}             & \multicolumn{1}{l}{0.6} & 20          & 0.022                & 10              \\
\multicolumn{1}{|c|}{DU(34, 56)} & \multicolumn{1}{c|}{}             & 0.7                     & 20          & 0.021                & 10              \\ \cline{2-6} 
\multicolumn{1}{|c|}{}           & \multicolumn{1}{c|}{}             & 0.3                     & 24          & 0.037                & 13              \\
\multicolumn{1}{|l|}{}           & \multicolumn{1}{l|}{}             & \multicolumn{1}{l}{0.4} & 25          & 0.036                & 13              \\
\multicolumn{1}{|c|}{}           & \multicolumn{1}{c|}{(0.05, 0.05)} & 0.5                     & 25          & 0.036                & 13              \\
\multicolumn{1}{|l|}{}           & \multicolumn{1}{l|}{}             & \multicolumn{1}{l}{0.6} & 26          & 0.036                & 13              \\
\multicolumn{1}{|c|}{}           & \multicolumn{1}{c|}{}             & 0.7                     & 27          & 0.038                & 13              \\ \hline
\multicolumn{1}{|c|}{}           & \multicolumn{1}{c|}{}             & 0.3                     & 20          & 0.022                & 11              \\
\multicolumn{1}{|l|}{}           & \multicolumn{1}{l|}{}             & \multicolumn{1}{l}{0.4} & 20          & 0.023                & 11              \\
\multicolumn{1}{|c|}{}           & \multicolumn{1}{c|}{(0.03, 0.03)} & 0.5                     & 21          & 0.022                & 11              \\
\multicolumn{1}{|l|}{}           & \multicolumn{1}{l|}{}             & \multicolumn{1}{l}{0.6} & 21          & 0.022                & 11              \\
\multicolumn{1}{|c|}{DU(10, 80)} & \multicolumn{1}{c|}{}             & 0.7                     & 22          & 0.023                & 11              \\ \cline{2-6} 
\multicolumn{1}{|c|}{}           & \multicolumn{1}{c|}{}             & 0.3                     & 27          & 0.035                & 14              \\
\multicolumn{1}{|l|}{}           & \multicolumn{1}{l|}{}             & \multicolumn{1}{l}{0.4} & 28          & 0.038                & 15              \\
\multicolumn{1}{|c|}{}           & \multicolumn{1}{c|}{(0.05, 0.05)} & 0.5                     & 28          & 0.035                & 14              \\
\multicolumn{1}{|l|}{}           & \multicolumn{1}{l|}{}             & \multicolumn{1}{l}{0.6} & 29          & 0.036                & 14              \\
\multicolumn{1}{|c|}{}           & \multicolumn{1}{c|}{}             & 0.7                     & 30          & 0.036                & 14              \\ \hline
\end{tabular}
\end{table}

\end{document}